# Plasmon triggered ultrafast operation of color centers in hBN layers

Vasilios Karanikolas[1,*], Takuya Iwasaki[2], Joel Henzie[2], Naoki Ikeda[3], Yusuke Yamauchi[2,4], Yutaka Wakayama[2], Takashi Kuroda[3], Kenji Watanabe[3], and Takashi Taniguchi[2]

1. International Center for Young Scientists (ICYS), National Institute for Materials Science, 1-1 Namiki, Tsukuba 305-0044, Japan
2. International Center for Materials Nanoarchitectonics (MANA), National Institute for Materials Science, 1-1 Namiki, Tsukuba 305-0044, Japan
3. Research Center for Functional Materials (RCFM), National Institute for Materials Science, 1-1 Namiki, Tsukuba 305-0044, Japan
4. School of Chemical Engineering and Australian Institute for Bioengineering and Nanotechnology, The University of Queensland, Brisbane, QLD, 4072 Australia

*KARANIKOLAS.Vasileios@nims.go.jp

## Abstract

High-quality emission centers in two-dimensional materials are promising components for future photonic and optoelectronic applications. Carbon-enriched hexagonal boron nitride (hBN:C) layers host atom-like color-center (CC) defects with strong and robust photoemission up to room temperature. Placing the hBN:C layers on top of Ag triangle nanoparticles (NPs) accelerate the decay of the CC defects down to 46 ps from their reference bulk value of 350 ps. The ultrafast decay is achieved due to the efficient excitation of the plasmon modes of the Ag NPs by the near field of the CCs. Simulations of the CCs/Ag NP interaction present that higher Purcell values are expected, although the measured decay of the CCs is limited by the instrument response. The influence of the NP thickness to the Purcell factor of the CCs is analyzed. The ultrafast operation of the CCs in hBN:C layers paves the way for their use in demanding applications, such as single-photon emitters and quantum devices.

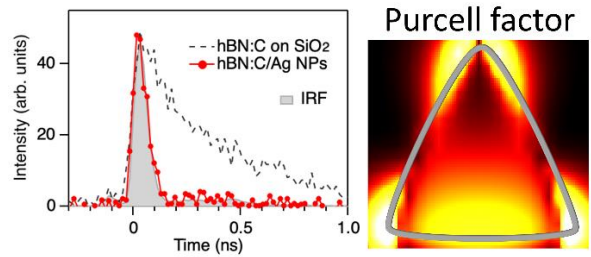

# Introduction

Engineering light-matter interactions at the nanoscale is essential for developing next-generation photonic and optoelectronic components for quantum computing, communication and sensing applications [1,2]. In recent years, two-dimensional (2D) materials have been explored experimentally and theoretically as nanophotonic building blocks because they possess better optical and mechanical properties than conventional bulk dielectric and noble metal materials [3,4]. For example, encapsulating 2D materials like graphene and transition metal dichalcogenide (TMD) monolayers in hexagonal boron nitride (hBN) layers increase their quality, boosting the field of nanoscience [5,6]. hBN is a material with unique optical properties, which include far-ultraviolet excitonic emission, hosting various localized color centers with emission in the visible wavelength region, as well as supporting phonon polariton modes [7]. Phonon polariton modes are hybrid modes of the electromagnetic (EM) field and the phonons of the hBN material, enabling hBN nanostructures to manipulate and focus light in the mid-infrared region of the EM spectrum [8,9].

hBN is an ultrawide band-gap (~6 eV) material [10,11] that can host myriad emission centers with different resonance wavelengths [12-15]. The exact nature and properties of the various defects are under theoretical investigation [16,17]. Emission centers in hBN have presented a plethora of phenomena in quantum optics, among them single-photon emission [18,19] with high quantum efficiency [20], and Rabi oscillations probed by the phonon sideband emission [21]. Moreover, controlled placement of color centers (CC) in hBN layers at specific positions has been demonstrated [22].

Localized photon sources can be hosted in uniform dielectric media. Their optical properties are measured via transient optical microspectroscopy with lifetimes typically spanning 1-10 ns. Due to their slow decay, the emitters become susceptible to various dephasing effects when they interact with their environment for long time spans [23]. Hence, the need to accelerate the decay of the emitters so as to operate at the quantum level.

To control and/or accelerate the decay of the defects in hBN layers they are coupled with dielectric [25-27] or metallic [28,29] cavities. The most widely used cavity are so-called "nanopatch antennae" composed of a noble metal nanocube on a noble metal film where CCs are hosted in hBN layers are placed between the particle and film [30,31]. The reference lifetime of the CCs is in the range of 1-5 ns and by placing them in the metallic cavity their radiative lifetime is accelerated to ~500 ps [32].

We focus on carbon-enriched hBN (hBN:C) [24], which possesses numerous optical emission peaks spanning the visible and near-infrared parts of the EM spectrum. Among the different emission centers of the hBN:C, the atom-like CC defect emission, with resonance wavelength at 804 nm, is significant because of its strong intensity up to room temperatures. The relaxation process of the atom-like CC has been investigated in bulk hBN:C crystals.

The CCs of the hBN:C bulk crystal are already high quality photon sources, with a lifetime of around 350 ps. The CC emission properties are maintained even when the material is mechanically exfoliated into thin 2D nanosheets, but their intensity is reduced due to the smaller volume.

In this manuscript, we present the accelerated operation of the atom-like CCs in carbon-enriched hBN (hBN:C) layers that are placed on top of triangular Ag nanoprisms (NPs). The hBN:C layers were initially transferred from the hBN:C crystal to a $SiO_2$/Si substrate to measure the intrinsic relaxation process of the target defects. Separately, Ag NPs were affixed to a different SiO2/Si substrate via the Langmuir Blodgett technique. Then the hBN:C layers were moved via a dry transfer technique and positioned on top of the Ag NPs for optical experiments. The accelerated decay of the CCs from 350 ps down to 46 ps is due to the excitation of the surface plasmon modes of the Ag NPs. From the lifetimes of the CCs on and off the Ag NPs a Purcell factor of 8 is extracted. By numerical simulations of the atom-like CC interaction with an Ag NP, high Purcell factor values are extracted that their values depend on the position of the CC with respect to the NP. The ultrafast operation of the atom-like CCs at elevated temperatures allows them to operate at the quantum level, opening compelling opportunities for developing ultra-compact quantum applications.

# Sample preparation, experimental and theoretical methods

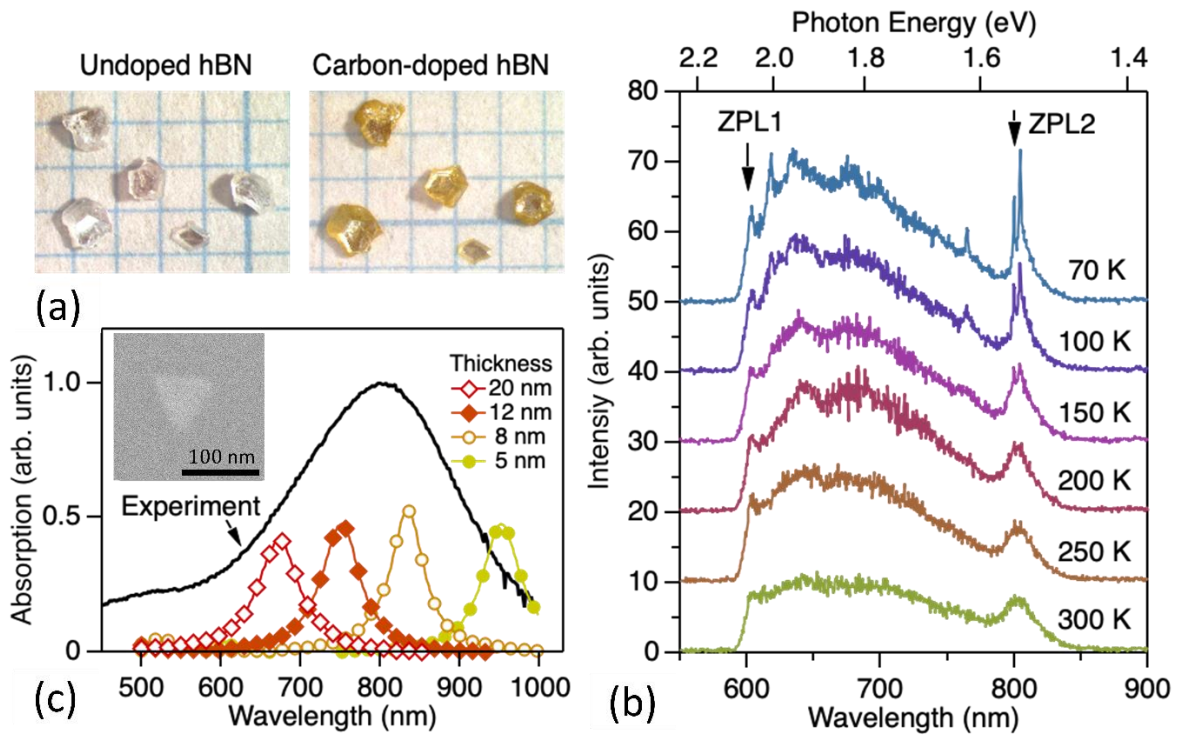

**Figure 1** (a) Photographs of the undoped hBN crystals (left) and the carbon-doped crystals prepared via high-temperature annealing (right). (b) Emission spectra of the hBN:C bulk crystal for different temperatures. The vertical arrows indicate the spectral position of the zero-phonon lines. (c) Extinction spectra of the Ag triangle nanoparticles in solution and simulated absorption spectra with varying thickness; (inset) SEM image of the nanoparticles when placed on the $SiO_2$/Si substrate.

## Sample preparation

Colorless and transparent hBN single crystals are grown via the temperature gradient method under high-pressure and high-temperature conditions [10]. The hBN crystals were annealed at 2,000 °C for 1 hour with nitrogen gas flow in a high-frequency furnace with a graphite susceptor. This process leads the hBN crystals to change their appearance from colorless and transparent to yellow, indicating that carbon impurities are incorporated in the hBN (see, Figure 1(a)). The doped crystals exhibit visible to near-infrared photoluminescence when excited with a laser. Figure 1(b) shows the luminescence spectra of the hBN:C bulk sample for different temperatures, with an excitation wavelength of 532 nm. The spectrum at 70 K consists mainly of two components. The shorter wavelength component has a sharp zero-phonon line at 604 nm (denoted by ZPL1) and a phonon-assisted broadband emission. The longer wavelength component has a strong zero-phonon peak at 804 nm (denoted by ZPL2) and lacks a significant phonon sideband emission. The zero-phonon peak at 804 nm has a doublet structure, which is possibly due to the different microscopic configurations of carbon impurities inside the crystal. Its behavior suggests a localized intra-defect transition, thus we refer to it as an atom-like CC defect. A similar spectral signature was observed previously in hBN [24]. Both the zero-phonon and atom-like resonant emission peaks are broadened with increasing temperature, although the 804 nm peak is evident even at room temperature. Hence, in this work we focus on the 804 nm luminescence state as the target localized photon emitter. It is noteworthy that ZPL2 has the linewidth of around 3-4 meV at low temperatures. It is much larger than the homogeneous radiative width of ~12 μeV for the lifetime of 350 ps, as descussed later. The major source for the line broadening is ascribed to coupling with low-frequency phonons in the crystal.

We used chemically synthesized Ag NPs nanoantennae, which are commercially available from Dai Nippon Toryo Co.Ltd. [33]. Figure 1(c) shows the extinction spectrum of the Ag NPs in water solution, where we observe a broad absorption peak that centers at 800 (±30) nm, which overlaps the emission wavelength of hBN:C (804 nm). The measured absorption peak is attributed to the localized plasmon mode, thus Ag NPs are an ideal model system to study the near-field coupling between the metal NPs and the hBN:C emitters. The inset in Figure 1(c) is a SEM image of a typical Ag NP, which indicates a triangular platelet shape with an average edge length of 90 nm and a thickness of 10 to 20 nm. It is known that, for metal NPs with such an asymmetric shape, the plasmon resonance wavelength is roughly proportional to the aspect ratio (edge length/ thickness) [34]. Thus, a major source for spectral broadening is considered to be NP shape inhomogeneity.

In Figure 1(c) we also present numerically simulated spectra of the Ag triangular NPs with different shapes, where we changed the thickness from 8 to 20 nm while keeping the edge length fixed at 90 nm. We find that the thinner NPs have a longer resonant wavelength, as expected, and the measured spectrum is well reproduced with a model NP ensemble thicknesses in the 10-20 nm provided by the manufacturer. Our simulation data verify that there is a distribution of particles in the solution.

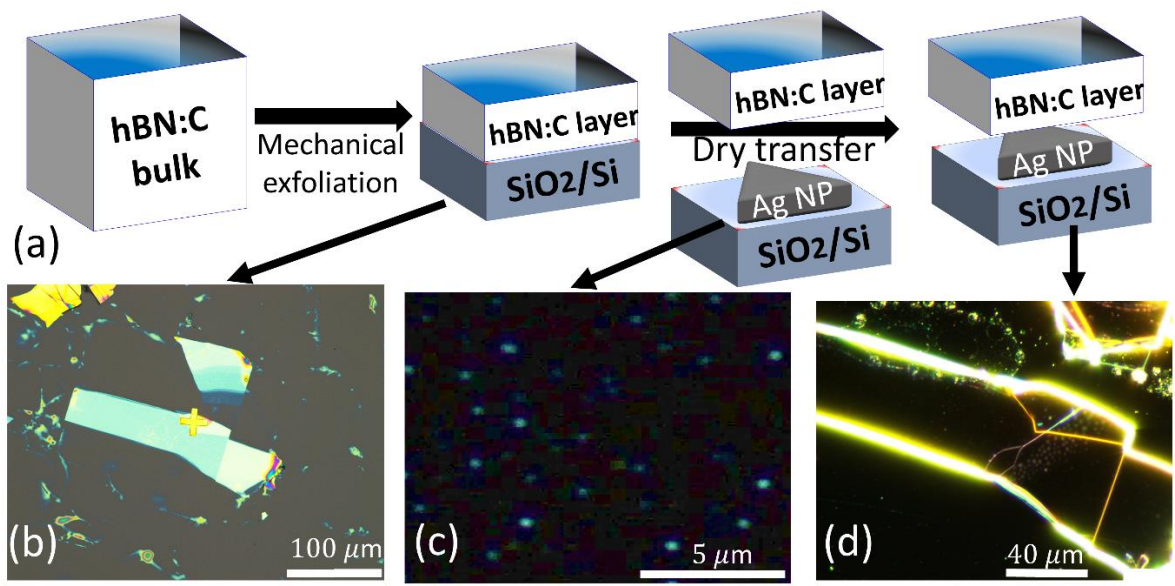

**Figure 2** (a) The different stages for preparing the hBN:C layer/Ag NPs/SiO$_2$/Si nanostructure. Optical microscope images of (b) the 26-nm thick hBN:C layer, (c) the Ag NPs and (d) of the transferred hBN:C layer on top of the Ag NPs..

The hBN flakes were transferred on the SiO$_2$/Si substrate by the mechanical exfoliation method using adhesive tape. The Ag NPs were dispersed on the SiO$_2$/Si substrate using an in-house Langmuir-Blodget trough [35]. A dry transfer of the hBN flakes onto the Ag NP was performed using the laboratory-built transfer system, detailed in [37]. The sample preparation is presented in Figure 2(a). In Figures 2(c-d), we present optical microscope images of the different stages for preparing the final sample. Numerous samples were created with this method, but we focus on the samples presented in Figure 2(b) using an hBN:C layer with a thickness of 26nm. The thickness of hBN was estimated to be 26 nm based on the optical contrast between the hBN and the substrate and later confirmed using atomic force microscopy, SIFigure 2 of [36]. In Figure 2(c) we present a dark field microscope image of the Ag NPs on the SiO$_2$/Si substrate. The hBN:C layer is placed on top of the Ag triangle NP through the dry-transfer technique that allow as to transfer the hBN:C layer at a specific place of interest, as presented in Figure 2(d), forming the hBN:C layer/ Ag triangle NPs/ SiO$_2$/Si structure we focus in this manuscript.

## Measurement setups

A home-built confocal setup was used to measure the photoemission signals. The excitation beam from a ps mode-locked Ti sapphire laser was focused on the sample using an objective lens with a numerical aperture of 0.55. The emission signal was collected by the same objective lens, coupled to a single-mode optical fiber, and fed into a spectrometer equipped with a charge-coupled device detector for spectral characterization and a microchannel plate photomultiplier tube (MCP-PMT) for time-resolved measurement [38]. The spectral window for the MCP-PMT was ~3 nm. Electric output from the MCP-PMT was sent to a time-correlated single-photon counter, which measured photon arrival with a time resolution of ~40 ps in full width at half-maximum (fwhm). All measurements were performed at 70 K.

## Computational methods

The atom-like CC defects in hBN:C layer are approached as two-level systems, and their Purcell factor is calculated through the macroscopic quantum electrodynamic (MQED) theory [39,40]. The excitation created by an electric dipole source is connected in quantum mechanical terms with the relaxation of the CC. The commercial finite difference time domain (FDTD) software from Ansys is used to solve the Maxwell equation considering an electric point dipole excitation in the presence of the Ag NP, see Figure 2(a). The optical response of the different materials is given by their experimentally measured dielectric permittivity data. Physically the Purcell factor value gives the enhancement or inhibition of the relaxation rate of the CC in the presence of the Ag triangle NP.

# Results and discussion

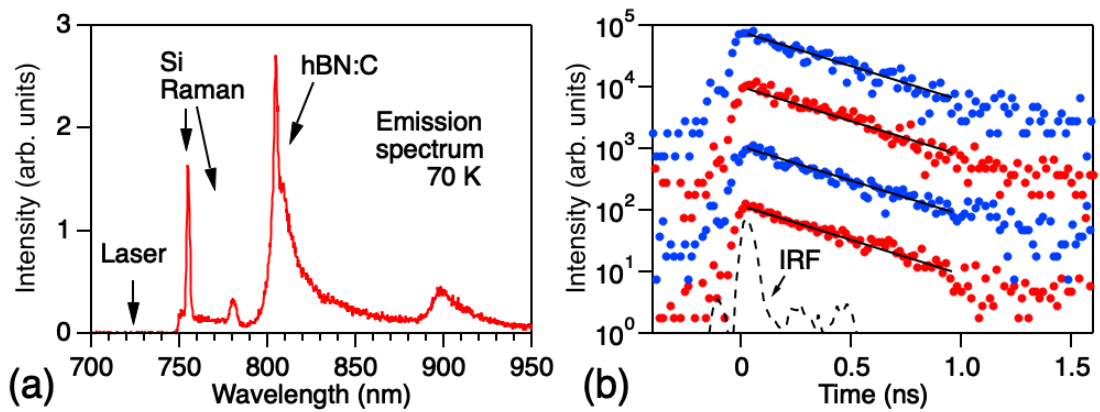

**Figure 3** (a) Typical emission spectrum of the exfoliated hBN:C film under phonon-assisted excitation at a wavelength of 724 nm. (b) Time-resolved emission signals of the color center emission at 804 nm for different hBN:C nanosheets on SiO$_2$. The solid lines are the single exponential fit with a decay time constant of 350 ps. The broken line shows the instrumental response function.

We start our analysis by measuring the emission spectrum of exfoliated hBN:C films on the SiO$_2$/Si substrate prior to placing them on the Ag NPs. We note that the emission intensity is strongly dependent on measured films and is generally much weaker than the bulk hBN:C (Figure 1(b)) due to the restriction in the excitation volume. To boost the signal intensity, we set the excitation wavelength to 724 nm, which allows us to excite the 804-nm defect state through a phonon-assisted transition (the TO phonon energy of hBN is 169.5 meV). The excitation spectrum reveals the significant resonant enhancement of the emitted signal, as shown in supplementary material (SIFig. 4). Figure 3(a) presents the spectrum of an hBN:C film (thickness > 150 nm) under the phonon-assisted excitation, where a long-pass filter with a cut-on wavelength of 750 nm was inserted to reject the intense scattered laser light. The two sharp peaks at 755 and 760 nm are attributed to Raman signals from the Si substrate. At the long-wavelength side of these Raman peaks, the CC defect emission from hBN:C is clearly observed at 804 nm due to the efficient excitation scheme.

Figure 3(b) is the semi-logarithmic plot of the emission decay signals at 804 nm for different exfoliated films. The intensities of the different curves are separated by a multiplication constant ($\times 10$) for clarity. All decay curves are well described by a single exponential expression $A\,exp(-t/\tau)$. The decay time constant $\tau$ is roughly independent of film thickness, although the absolute intensity is quite different; see SIFigure 3 of the SI for the unnormalized intensity plots [36]. These observations imply that every CC dispersed in the hBN:C layer sample has the same decay dynamics, governed by spontaneous radiative emission that is free from nonradiative relaxations. Through fitting, we extract the radiative lifetime to be $\tau_{ref} = 350\ (\pm 20)$ ps. The experimental results presented at Figure 3(b) were obtained at a temperature of 70 K. In the supplementary materials we present that at elevated temperatures, up to room temperature (300 K), the measured lifetime of the CCs remains almost unchanged (See, SIFig. 5). These results emphasize that the atom-like CC defects of the hBN:C nanosheet are extremely high quality quantum photon sources, well protected in the hBN:C lattice. Thus, the intrinsic non-radiative losses are small, showing that the color-centers in hBN:C have a high quantum efficiency.

Using the atomic description for the lifetime, $\tau_{ref} = 3\pi c^3 \hbar \varepsilon_0 \lambda^3/[8\pi^3 \mu^2]$, we obtain a transition dipole moment of $\mu = 42$ D for the atom-like CC in hBN:C. Such a large moment value has also been observed by excitons in transition metal dichalcogenides (TMDs) like WS$_2$; although TMD systems have limitations for their application to light-emitting devices [41]. For example, the intense photoluminescence of WS$_2$ monolayers is maintained by encapsulating them within a hBN layer. On the other hand, hBN:C layers have their own high emission yield and high-operation rate quantum emitters and can be directly and precisely transferred to a target area.

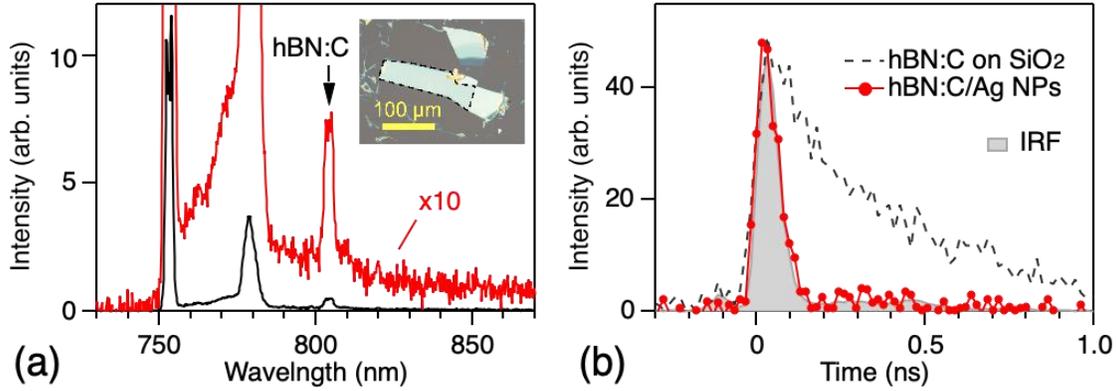

**Figure 4** (a) Emission spectrum of the ultrathin hBN:C layer interacting with the Ag NPs. The red line is the expanded curve (x10). The inset is the optical microscope image of the measured hBN:C sample. (b) Comparison between the emission decay curves for the hBN:C layer on the Ag NPs (red circles) and the SiO2 substrate (broken line). They are normalized to their peak intensity. The gray shade indicates the instrumental response function.

We now investigate the impact of the metal NPs on the atom-like CCs in hBN:C. To elucidate the near-field response of the CCs, whose characteristic length is down to a few nanometers, we prepared relatively thin hBN:C films, which were then transferred on the dispersed Ag triangle NPs on the SiO$_2$/Si substrate, as shown in Figure 2(d). The thickness of our target film is 26 ($\pm$1) nm according to atomic force microscopy (AFM), SIFigure 2 of [36]. Despite the small thickness, we can identify significant emission signals from the target atom-like CC at 804 nm (Fig. 4(a)). Furthermore, the position-dependent measurement results confirm that the detected emission comes from the CCs of the hBN:C film, as shown in supplementary material (SIFig. 6).

In Figure 4(b) we present the emission decay curve of the hBN:C ultrathin film placed on top of the Ag NPs, together with that of hBN:C films alone for comparison, where the latter curve is given by the average of the decay signals shown in Figure 3(b). Remarkably, the hBN:C/Ag NP composite exhibits a very fast decay, which roughly follows the instrumental response function (IRF) of our setup. The single exponential fitting to the measured curve reveals the decay lifetime constant $\tau_{NP} = 46$ ($\pm$10) ps, which is shorter by a factor of around 8 compared with that of the hBN:C film alone ($\tau_{ref} = 350$ ps). Thus, the experimentally obtained Purcell factor, $\Gamma_{exp}$, is given by $\Gamma_{exp} = \tau_{ref}/\tau_{NP} \sim 8$. Since the IRF is close to the measured curve, the actual decay is expected to be even faster. It is worth noting that such lifetime shortening is not simply due to the emergence of internal nonradiative relaxation channels, potentially associated within the composite sample, because this interpretation is contradicted by the presence of the defects' emission spectrum profile in Figure 4(a). Instead, the measured fast decay is expected to be due to the near-field coupling between the CCs and Ag NPs, which leads to the substantially accelerated photoemission, as will be supported by the theoretical analysis below.

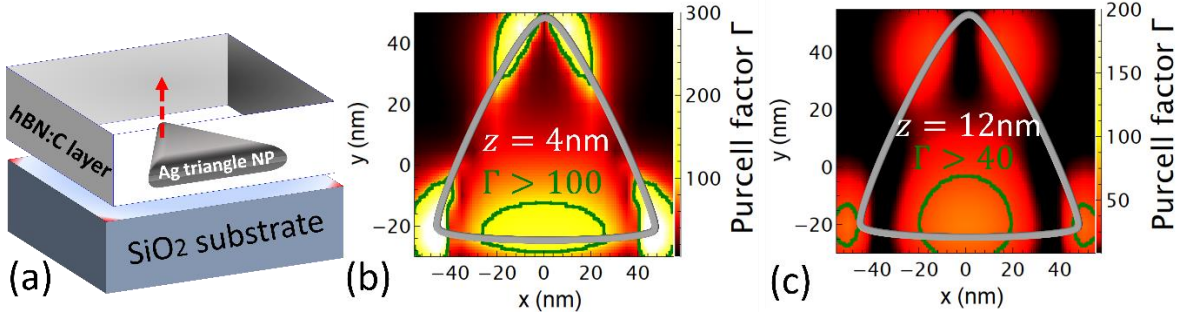

**Figure 5** (a) The simulated nanostructure used to study the atom-like CCs interacting with an Ag NP, where we consider an hBN layer with a thickness of 26 nm and a triangular NP with an edge length of 90 nm and a thickness of 16 nm. (b and c) Contour plots of the Purcell factor on the x-y plane, which is positioned (b) 4 nm and (c) 12nm above an Ag NP. The silver line indicates the triangular edge of the NP. The transition dipole moment of the CC is along the x-axis and the emission wavelength is 804 nm.

Numerical simulations are used to clarify the near-field interaction between the atom-like CCs and the Ag NPs. Figure 5(a) shows the structure used to study the CC emission. We focus on the interaction of the CCs with an isolated triangular Ag NP that has an edge length of 90 nm and a thickness of 16 nm. Then, the Ag NP is placed between a 26-nm-thick hBN:C layer and 90-nm-thick $SiO_2$, the parameters of which resemble the experimental condition.

We estimate there are ~$10^2$ defects above the triangular area of a single NP based on the defect density of $\sim 2.4 \times 10^{-4} nm^{-2}$ in an hBN:C monolayer, which has been measured using the STM technique [42]. Hence, we deal with emission signals from several CCs at different positions with respect to the Ag NP. To address the effect of the emitter distribution, we analyze the Purcell factor $\Gamma(\mathbf{r}_{CC}, \lambda)$ as functions of the CC position $\mathbf{r}_{CC}$, the emission wavelength $\lambda$, and the thickness of the Ag NP. Then, the emission lifetime of the CC defect is given by $\tau'_{NP} = \tau_{ref} / \Gamma(\mathbf{r}_{CC}, \lambda)$, and the emission decay signal is expected to follow an exponential decay, i.e., $\exp(-t/\tau'_{NP}) = \exp(-\Gamma(\mathbf{r}_{CC}, \lambda)t/\tau_{ref})$, accelareted from the $\tau_{ref}$ value by a factor of $\Gamma$.

In Figure 5(b) we present a contour plot of the Purcell factor $\Gamma$ of a single CC, placed 4 nm above the Ag NP in the z-direction, scanning the x-y plane. Here, we assume that the CC emitter has a transition dipole moment along the x-axis and an emission wavelength of 804 nm. The red lines highlight the region where the Purcell factor of the CC has values higher than 100, translating to $\tau'_{NP} < 3.5$ ps. Due to the x transition dipole moment orientation of the CC, the dipole plasmon mode of the Ag NP is excited, leading to the highest enhancement values along the edge of the triangular NP. Such strong Purcell enhancement is also related to the short vertical distance between the CC emitters and the Ag NP ($z_{CC} = 4$ nm). The total Purcell factor value $\Gamma$ has two contributions the radiative $\Gamma_r$ and the nonradiative $\Gamma_{nr}$. The experimentally measured counts come from the radiative emission of the CCs. The near-field of the CC, with emission wavelength of 804nm, excites the dipole plasmon mode supported by the Ag triangular NP, this mode is used here to transfer the near-field excitation to the far-field. In the Supplementary Information we present that in the encircled area of Figure 5(b) the radiative Purcell factor has values $\Gamma_r > 25$, while the quantum efficiency reach values $\Gamma_r/\Gamma > 40\%$. This information further supports that the experimentally measured counts to extract the lifetime in the case of the CCs interact with the Ag nanoparticle are of the radiative emission, and the nonradiative contribution cannot be detected due to the limitation by the instrument response.

The strong Purcell factor enhancement effect of the CC remains even for longer distances, Figure 5(c) shows a contour plot of the CC Purcell factor for $z_{CC} = 12$ nm, close to the middle of the hBN:C layer. It reveals significant Purcell enhancement, which reaches values $\Gamma = 40$ indicated by the green lines. We observe that the largest enclosed area is in the middle of the Ag NP edge due to the efficient excitation of the dipole plasmon mode that has the highest extend in the perpendicular direction compared with the higher order modes. On the other hand, when the CC is closer to the Ag edges, the Purcell factor value drops faster as the separation distance increases. For the $z_{CC} = 12$nm, the spatially averaged Purcell factor to be 19 with a standard deviation of 15, which results in $\tau'_{NP} = 18\ (\pm15)$ ps.

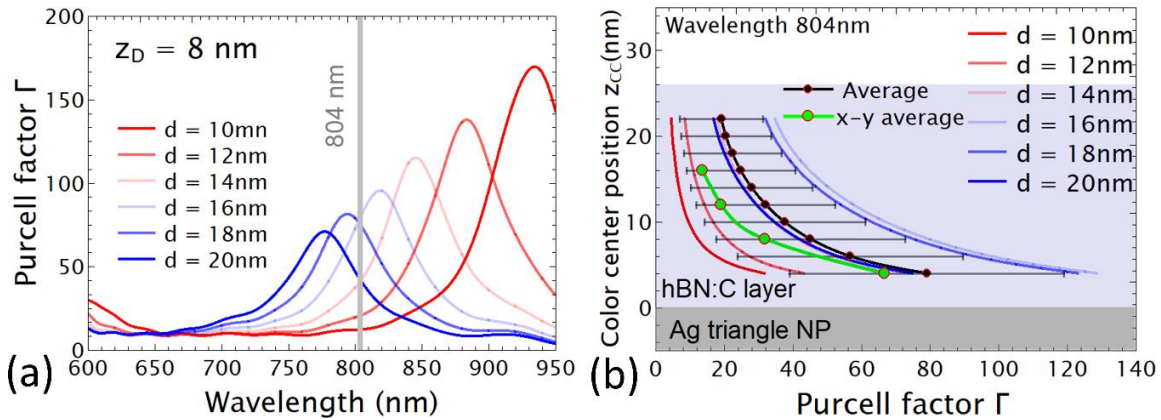

**Figure 6**. (a) The Purcell factor of an hBN:C CC with a fixed position and varying the emission wavelength. (b) The Purcell factor of an hBN CC with a fixed emission wavelength of $\lambda = 804$ nm and varying the CC/NP separation distance. In (a,b) different Ag NP thicknesses are considered. The transition dipole moment orientation of the CC is along the x-axis.

We also study the impact of the NP shape variation on the emission dynamics. In Figure 6(a) we present the Purcell factor spectrum for the Ag NPs with different thicknesses and a constant edge length (90 nm). In this analysis, we monitor the CC placed at a fixed position, namely, in the middle above the long edge of the triangular NP and $z_D = 8$ nm. The plasmon resonance wavelength redshifts as the Ag NPs become thinner, following a similar trend as observed in Figure 1(c). For the CC with an emission wavelength of 804 nm, the highest Purcell factor value is ~90 and is achieved for Ag NP thicknesses between 16 to 18 nm.

In Figure 6(b) we present the vertical position dependence of the Purcell factor for the Ag NPs with different thicknesses. The emitter position follows the red arrow in Figure 5(a) and it emits at a wavelength of 804 nm. We observe that as the CC/Ag NP separation is reduced, the Purcell factor value increases. Moreover, for the NP thicknesses between 16 to 18nm, the Purcell factor value has the largest value, according to Figure 6(a). We also plot the mean of the Purcell factor values for the NPs with different thicknesses, and the standard deviation values are also included as error bars. The result implies that the significant Purcell effect with Γ~20 occurs even when the CC is far from the Ag NP and close to the top surface of the sample layer.

In the experiment, we extracted Purcell factors as large as $\Gamma_{exp}$ ~8, which is probably a conservative estimate because our measurement is limited by the instrumental response. Hence, our samples are estimated to possess higher Purcell values, as predicted by the above theoretical analysis. We also analyze the experimental decay profile in Figure 4(b) and reveal that 29% of the CCs achieve Purcell factor values above $\Gamma_{exp} = 8$, see Supporting Figure 8(b) [43-45]. Consequently, we successfully demonstrate the ultrafast photoemission from the hBN CCs in the nearfield region of the Ag NPs, both experimentally and theoretically.

In this manuscript, we focus on a hBN:C layer with fixed thickness of 26nm and theoretically investigate the effect of varying the Ag triangle NP thickness. Increasing the hBN:C layer thickness will also redshift the plasmonic resonance of the Ag NP. The effect is similar to the case of noble metal NPs coated by a dielectric material, where increasing the coating thickness redshifts to the plasmonic resonance up to some thickness value that resembles the homogeneous space [46].

# Conclusions

The decay of the atom-like CC defects of hBN:C layers is engineered by placing them close to triangular Ag NPs. The hBN:C layers contain high-quality CC emitters with small internal losses. The near field of the CCs excites the plasmonic modes supported by Ag NP, accelerating relaxation of the CC excited state from 350 ps to 46 ps when the CCs interact with the Ag NP. This resulted in a Purcell value of $\Gamma_{exp}$~8 for the Ag NP coupled hBN:C CCs. Higher Purcell factors values are expected, although the measured decay of the CCs is limited by the instrument response. Numerical simulations also indicate that higher Purcell values can be achieved experimentally, especially when the CCs are close to the edges of the Ag triangle NPs. The thickness distribution of the Ag NPs affects the Purcell factor of the CC. The ultra-fast operation of the CCs in hBN:C layers can reduce dephasing effects, an important feature for quantum applications.

# Aknowledgements

V.K. research was supported by JSPS KAKENHI Grant No. JP21K13868. K.W. research was supported by JSPS KAKENHI Grant No. JP22H01975.